\documentclass[
    ,final            % use final for the camera ready runs
  ]
  {aipproc}

\layoutstyle{8x11single}

\usepackage{amsmath}
\usepackage{amsfonts,amsbsy,amsxtra}
\usepackage{amssymb,latexsym}
\usepackage{textcomp}
\usepackage{ifthen}
\usepackage{lineno}
\usepackage{blindtext}

\newcommand{\figref}[1]{{Fig.~\ref{fig:#1}}}
\newcommand{\Figref}[1]{{Figure~\ref{fig:#1}}}

\newcommand{\others}{\textit{et al.}}

\newcommand{\antibar}[1]{%
  \overline{#1}%
}
\newcommand{\PPh}{\ensuremath{\gamma}}

\renewcommand{\Pr}{\ensuremath{\rho}}
\newcommand{\Prthree}{\ensuremath{\rho_3}}

\newcommand{\Paone}{\ensuremath{a_1}}
\newcommand{\Patwo}{\ensuremath{a_2}}
\newcommand{\Pafour}{\ensuremath{a_4}}

\newcommand{\Pfzero}{\ensuremath{f_0}}

\newcommand{\Pftwo}{\ensuremath{f_2}}

\newcommand{\Ppi}{\ensuremath{\pi}}
\newcommand{\Ppione}{\ensuremath{\pi_1}}
\newcommand{\Ppitwo}{\ensuremath{\pi_2}}
\newcommand{\Ppiz}{\ensuremath{\pi^0}}
\newcommand{\Ppip}{\ensuremath{\pi^+}}
\newcommand{\Ppim}{\ensuremath{\pi^-}}

\newcommand{\Peta}{\ensuremath{\eta}}
\newcommand{\Petaprime}{\ensuremath{\eta'}}
\newcommand{\PK}{\ensuremath{K}}

\newcommand{\PKm}{\ensuremath{K^-}}

\newcommand{\PKstar}{\ensuremath{K^\ast}}
\newcommand{\PKone}{\ensuremath{K_1}}
\newcommand{\PKtwo}{\ensuremath{K_2}}
\newcommand{\PKtwostar}{\ensuremath{K_2^\ast}}

\newcommand{\Ppbar}{\ensuremath{\antibar{p}}}

\newcommand{\twopion}{\ensuremath{\Ppip\Ppim}}
\newcommand{\threepion}{\ensuremath{\Ppim\Ppip\Ppim}}

\newcommand{\abs}[1]{{|{#1}|}}

\newcommand{\measresult}[4]{%
  \ensuremath{#1%
    \ifthenelse{\equal{#2}{}}%
    {}%
    {\pm #2%
      \ifthenelse{\equal{#3}{}}%
      {}%
      {_\text{stat.}}%
    }%
    \ifthenelse{\equal{#3}{}}%
    {}%
    {\pm #3_\text{syst.}}\text{#4}%
  }%
}

\newcommand{\jpc}{\ensuremath{J^{PC}}}

\newcommand{\wavespec}[7]{\ensuremath{#1^{#2#3}\, #4^{#5}\,%
    \ifthenelse{\equal{#6}{}}
    {}
    {[{#6}]}
    #7}
}

\newcommand{\gevc}{~\ensuremath{\text{GeV}\! / c}}
\newcommand{\gevcsq}{~\ensuremath{(\text{GeV}\! / c)^2}}
\newcommand{\mevcc}{~\ensuremath{\text{MeV}\! / c^2}}
\newcommand{\gevcc}{~\ensuremath{\text{GeV}\! / c^2}}
\newcommand{\tenpow}[2][]{%
  \ifthenelse{\equal{#1}{}}
  {\ensuremath{10^{#2}}}
  {\ensuremath{{#1} \cdot 10^{#2}}}
}

\begin{document}

\title{Hadron Spectroscopy with COMPASS}


\author{Boris Grube on behalf of the COMPASS Collaboration}{
  address={Physik-Department E18, Technische Universit\"at M\"unchen,
    James-Franck-Str., D-85748 Garching, Germany}}

\begin{abstract}
  COMPASS is a multi-purpose fixed-target experiment at the CERN Super
  Proton Synchrotron aimed at studying the structure and spectrum of
  hadrons. One primary goal is the search for new hadronic states, in
  particular spin-exotic mesons and glueballs.
  %, using hadron beams.
  We present recent results of partial-wave analyses of $(3\Ppi)^-$
  and $\Ppim\Petaprime$ final states based on a large data set of
  diffractive dissociation of a 190\gevc\ \Ppim\ beam on a proton
  target in the squared four-momentum-transfer range $0.1 < t' <
  1\gevcsq$.
  %This reaction provides clean access to the light-quark
  %meson spectrum up to masses of 2.5~\gevcc.
  We also show first results from a partial-wave analysis of
  diffractive dissociation of $\PKm$ into $\PKm\Ppip\Ppim$ final
  states are presented.
\end{abstract}

\maketitle

%%%%%%%%%%%%%%%%%%%%%%%%%%%%%%%%%%%%%%%%%%%%
%% MAINMATTER
%%%%%%%%%%%%%%%%%%%%%%%%%%%%%%%%%%%%%%%%%%%%

\section{Introduction}

The COmmon Muon and Proton Apparatus for Structure and Spectroscopy
(COMPASS)~\cite{compass} is a fixed-target experiment at the CERN
Super Proton Synchrotron (SPS). It is a two-stage high-resolution
spectrometer that covers a wide range of scattering angles and
particle momenta. The spectrometer is equipped with hadronic and
electromagnetic calorimeters so that final states with charged as well
as neutral particles can be reconstructed. A Ring Imaging Cherenkov
Detector (RICH) in the first stage can be used for particle
identification. It is able to separate kaons from pions up to momenta
of 50\gevc. The target is surrounded by a Recoil Proton Detector (RPD)
that measures the time of flight of the recoil protons using two
scintillator barrels. COMPASS is connected to the M2 beam line of the
SPS which can deliver secondary hadron beams with a momentum of up to
300\gevc\ and a maximum intensity of
\tenpow[5]{7}~$\text{s}^{-1}$. The negative hadron beam that was used
for the analyses presented here has a momentum of 190\gevc\ and
consists of 96.8~\% \Ppim, 2.4~\% \PKm, and 0.8~\% \Ppbar\ at the
target. Two ChErenkov Differential counters with Achromatic Ring focus
(CEDAR) upstream of the target are used to identify the incoming beam
particles.

Diffractive dissociation reactions are known to exhibit a rich
spectrum of produced states. In these events the beam hadron is
excited to some intermediate state $X$ via $t$-channel Reggeon
exchange with the target. At 190\gevc\ beam energy Pomeron exchange is
dominant. The $X$ then decays into a $n$-body final state. The
process
\begin{equation}
  \text{beam} + \text{target} \to X + \text{recoil}, \qquad X \to h_1 \ldots h_n
\end{equation}
is characterized by two kinematic variables: the square of the total
center-of-mass energy, $s$, and the squared four-momentum transfer
%from the incoming beam particle
to the target, $t = (p_\text{beam} - p_X)^2$. It is customary to use
the variable $t' \equiv \abs{t} - \abs{t}_\text{min}$ instead of $t$,
where $\abs{t}_\text{min}$ is the minimum value of $\abs{t}$ for a
given invariant mass of $X$.

\section{$(3\Ppi)^-$ Final State from \Ppim\ Diffraction}

In a partial-wave analysis (PWA) of the pilot run data taken in 2004,
a significant spin-exotic $J^{PC} = 1^{-+}$ resonance was found at
around 1660\mevcc\ in $\pi^-\pi^+\pi^-$ final states produced in
\Ppim\ diffraction on a Pb target at squared four-momentum transfers
of $0.1 < t' < 1.0\gevcsq$~\cite{compass_exotic}. The resonance
parameters are consistent with the disputed $\Ppione(1600)$ claimed in
this channel by other experiments~\cite{exotic}.

In 2008 COMPASS has acquired large data sets of diffractive
dissociation of 190\gevc\ \Ppim\ on a $\ell$H$_2$ target. The trigger
included a beam definition and the RPD, which ensured that the target
proton stayed intact and also introduced a lower bound for $t'$ of
about 0.1\gevcsq. Events with charged particle trajectories outside
the spectrometer acceptance and those, where the beam particle
traversed the target unscattered, were vetoed. In the offline data
selection events were required to have a well-defined primary
interaction vertex inside the target volume. Diffractive events were
enriched by an exclusivity cut around the nominal beam energy. After
all cuts the \threepion\ sample from the 2008 run contains
\tenpow[96]{6}~events.

In the PWA the isobar model~\cite{isobar} is used to decompose the
decay $X^- \to \threepion$ into a chain of successive two-body
decays. The $X^-$ with quantum numbers \jpc\ and spin projection
$M^\epsilon$ is assumed to decay into a di-pion resonance, the
so-called isobar, and a bachelor pion. The isobar has spin $S$ and a
relative orbital angular momentum $L$ with respect to
$\Ppim_\text{bachelor}$. A partial wave is thus defined by $\jpc
M^\epsilon[\text{isobar}]L$, where $\epsilon = \pm 1$ is the
reflectivity~\cite{reflectivity}.

The spin-density matrix is determined by extended maximum likelihood
fits performed in 20\mevcc\ wide bins of the three-pion invariant mass
$m_X$. In these fits no assumption is made on the produced resonances
$X^-$ other than that their production strengths are constant within a
$m_X$ bin. The PWA model includes five \twopion\
isobars~\cite{compass_exotic}: $(\Ppi\Ppi)_\text{$S$-wave}$,
$\Pr(770)$, $\Pfzero(980)$, $\Pftwo(1270)$, and $\Prthree(1690)$. They
were described using relativistic Breit-Wigner line shape functions
including Blatt-Weisskopf barrier penetration
factors~\cite{bwFactor}. For the \twopion\ $S$-wave we use the
parametrization from~\cite{vesSigma} with the $\Pfzero(980)$
subtracted from the elastic $\Ppi\Ppi$ amplitude and added as a
separate Breit-Wigner resonance. In total the wave set consists of 52
waves plus an incoherent isotropic background wave. Mostly positive
reflectivity waves are needed to describe the data which corresponds
to production with natural parity exchange. A rank-two spin-density
matrix was used in order to account for spin-flip and spin-non-flip
amplitudes at the target vertex.

\begin{figure}[t]
  \includegraphics[width=\textwidth]{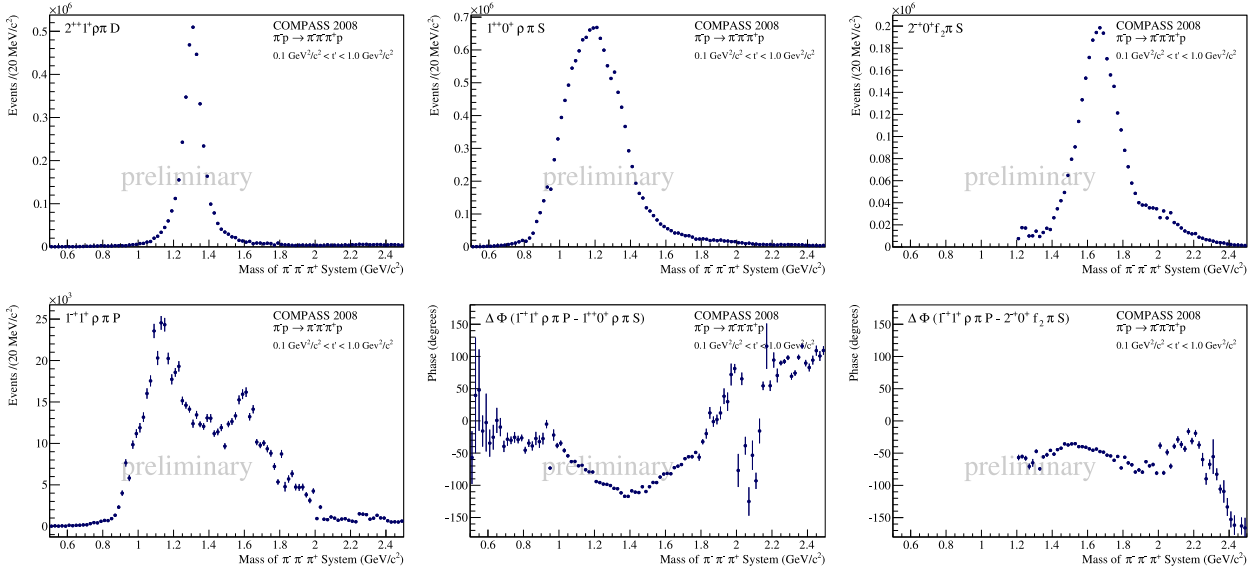}
  \label{fig:2008_3pi}
  \caption{\emph{Top row:} Intensities of major waves in \threepion\
    final state. \protect \wavespec{2}{+}{+}{1}{+}{\Pr\Ppi}{D} wave
    with $\Patwo(1320)$ (left), \protect
    \wavespec{1}{+}{+}{0}{+}{\Pr\Ppi}{S} wave with $\Paone(1260)$
    (center), and \protect \wavespec{2}{-}{+}{0}{+}{\Pftwo\Ppi}{S}
    wave with $\Ppitwo(1670)$ (right). \emph{Bottom row:} Intensity of
    the spin-exotic \protect \wavespec{1}{-}{+}{1}{+}{\Pr\Ppi}{P} wave
    (left) and its phase relative to the \protect
    \wavespec{1}{+}{+}{0}{+}{\Pr\Ppi}{S} (center) and the \protect
    \wavespec{2}{-}{+}{0}{+}{\Pftwo\Ppi}{S} waves
    (right). (From~\cite{hadron2011_florian})}
\end{figure}

The intensity of the three dominant waves in the \threepion\ final
state, \wavespec{1}{+}{+}{0}{+}{\Pr\Ppi}{S},
\wavespec{2}{+}{+}{1}{+}{\Pr\Ppi}{D}, and
\wavespec{2}{-}{+}{0}{+}{\Pftwo\Ppi}{S}, are shown in
\figref{2008_3pi}, top row. They contain resonant structures that
correspond to the $\Paone(1260)$, $\Patwo(1320)$, and $\Ppitwo(1670)$,
respectively~\cite{hadron2011_florian}. \Figref{2008_3pi} bottom, left
shows the intensity of the spin-exotic
\wavespec{1}{-}{+}{1}{+}{\Pr\Ppi}{P} wave. The plot nicely illustrates
the unprecedented statistical accuracy due to the large data set. This
wave exhibits a peak structure around 1.6\gevcc. In this mass region a
rising phase with respect to the tail of the $\Paone(1260)$ in the
\wavespec{1}{+}{+}{0}{+}{\Pr\Ppi}{S} wave is seen
(cf. \figref{2008_3pi} bottom, center). As \figref{2008_3pi} bottom,
right shows, the structure is phase locked with the $\Ppitwo(1670)$ in
the \wavespec{2}{-}{+}{0}{+}{\Pftwo\Ppi}{S} wave. This is consistent
with the results obtained from a PWA of the pilot-run data taken with
a Pb target~\cite{compass_exotic}. The bump around 1.2\gevcc\ is still
being investigated. It is unstable with respect to changes in the PWA
model which hints that it might be an artifact of the analysis
method. More detailed studies as well as a mass-dependent fit of the
spin-density matrix are under way.

\begin{figure}
  \parbox{0.25\textwidth}{\includegraphics[width=0.25\textwidth]{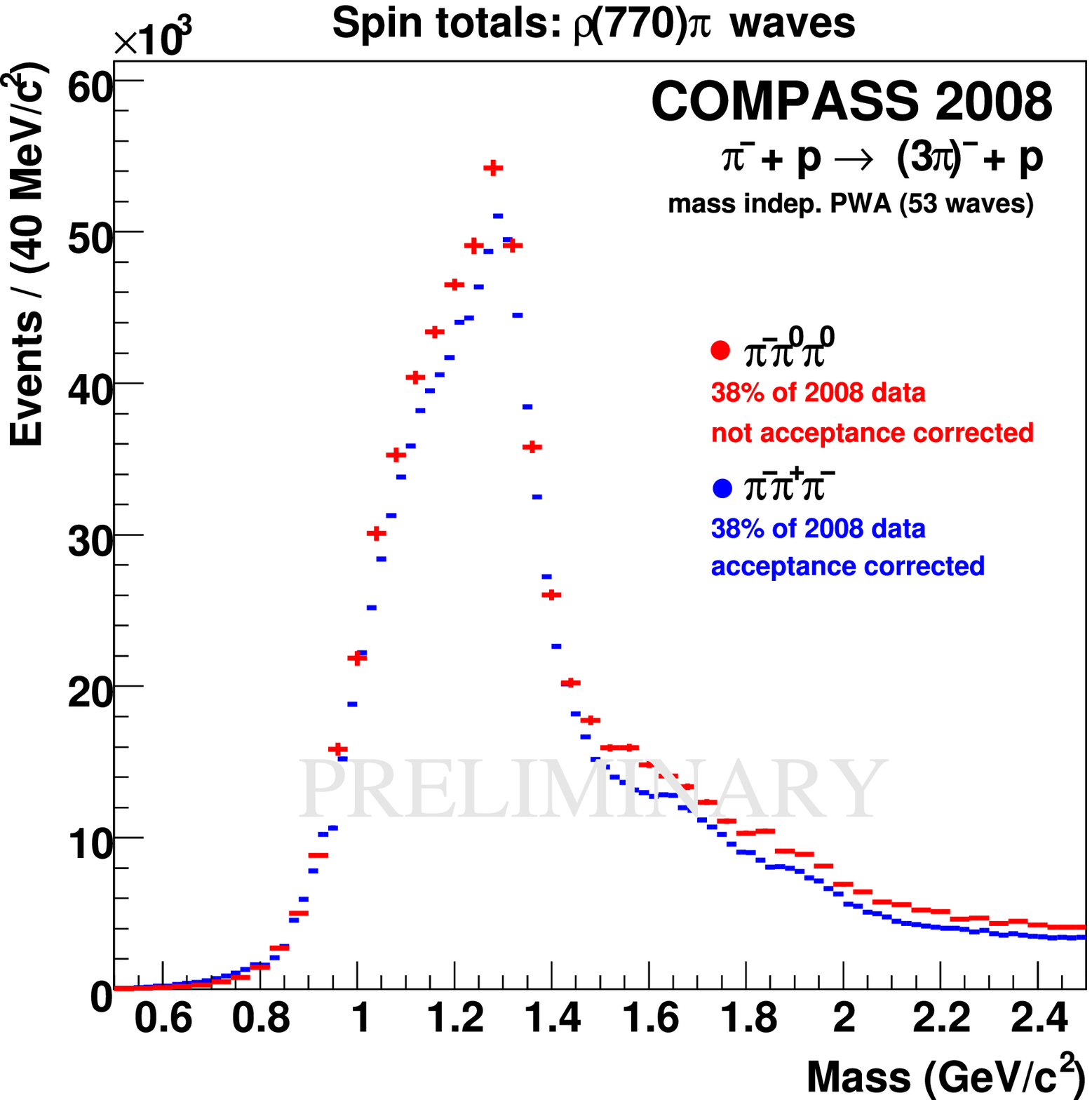} \\[-1.5ex] \centering (a)}
  \parbox{0.25\textwidth}{\includegraphics[width=0.25\textwidth]{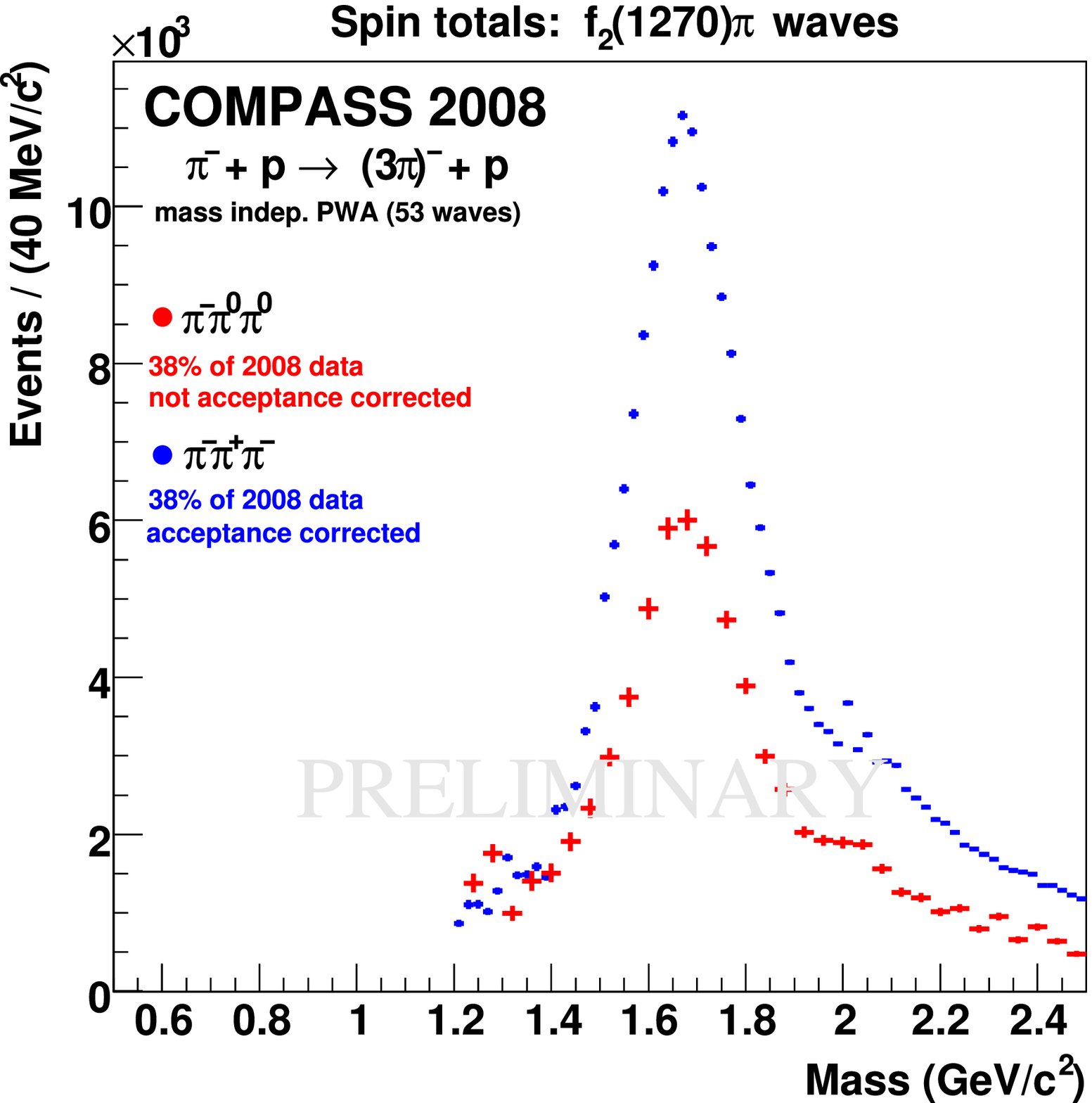} \\[-1.5ex] \centering (b)}
  \parbox{0.25\textwidth}{\includegraphics[width=0.25\textwidth]{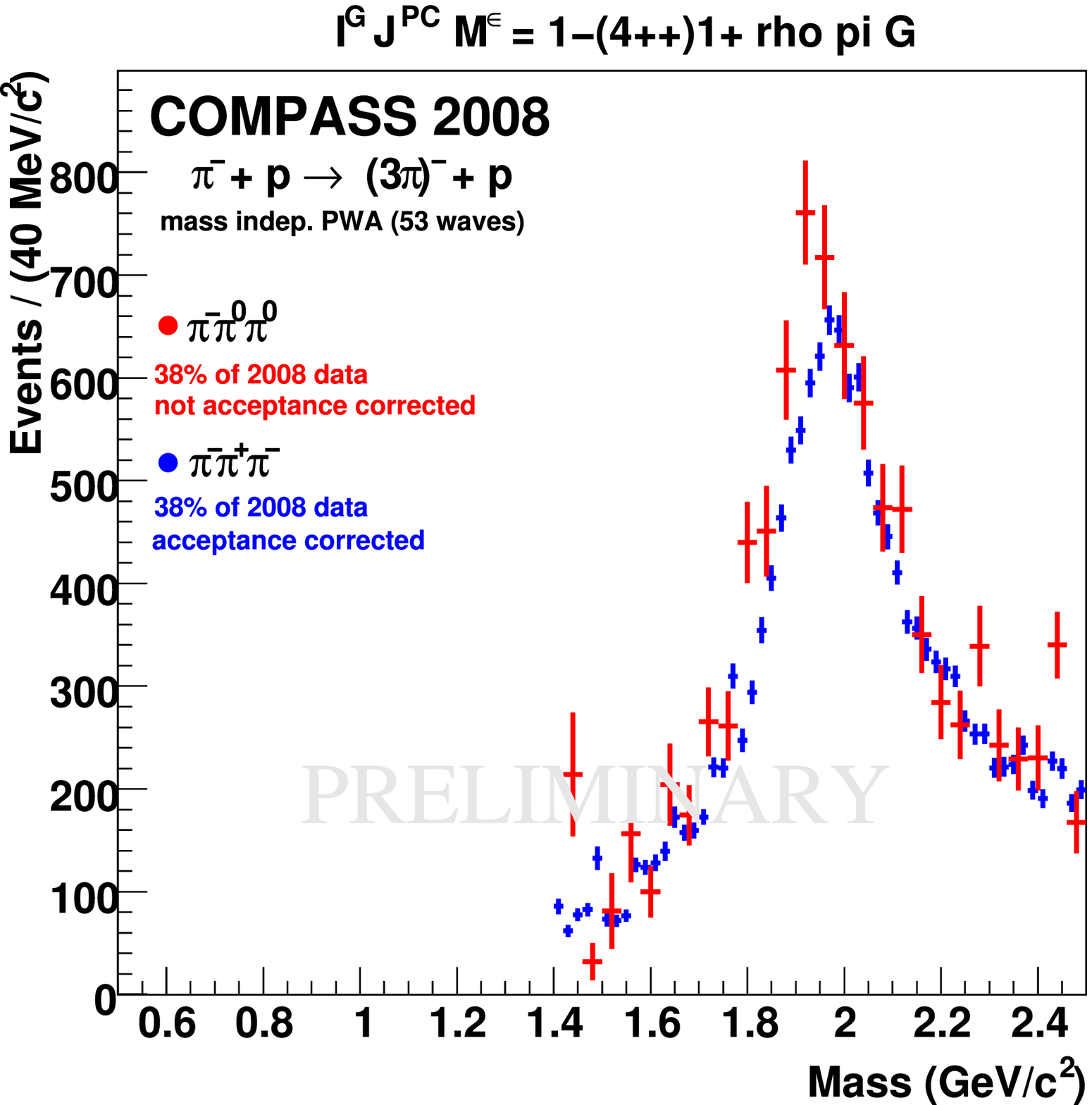} \\[-1.5ex] \centering (c)}
  \parbox{0.25\textwidth}{\includegraphics[width=0.25\textwidth]{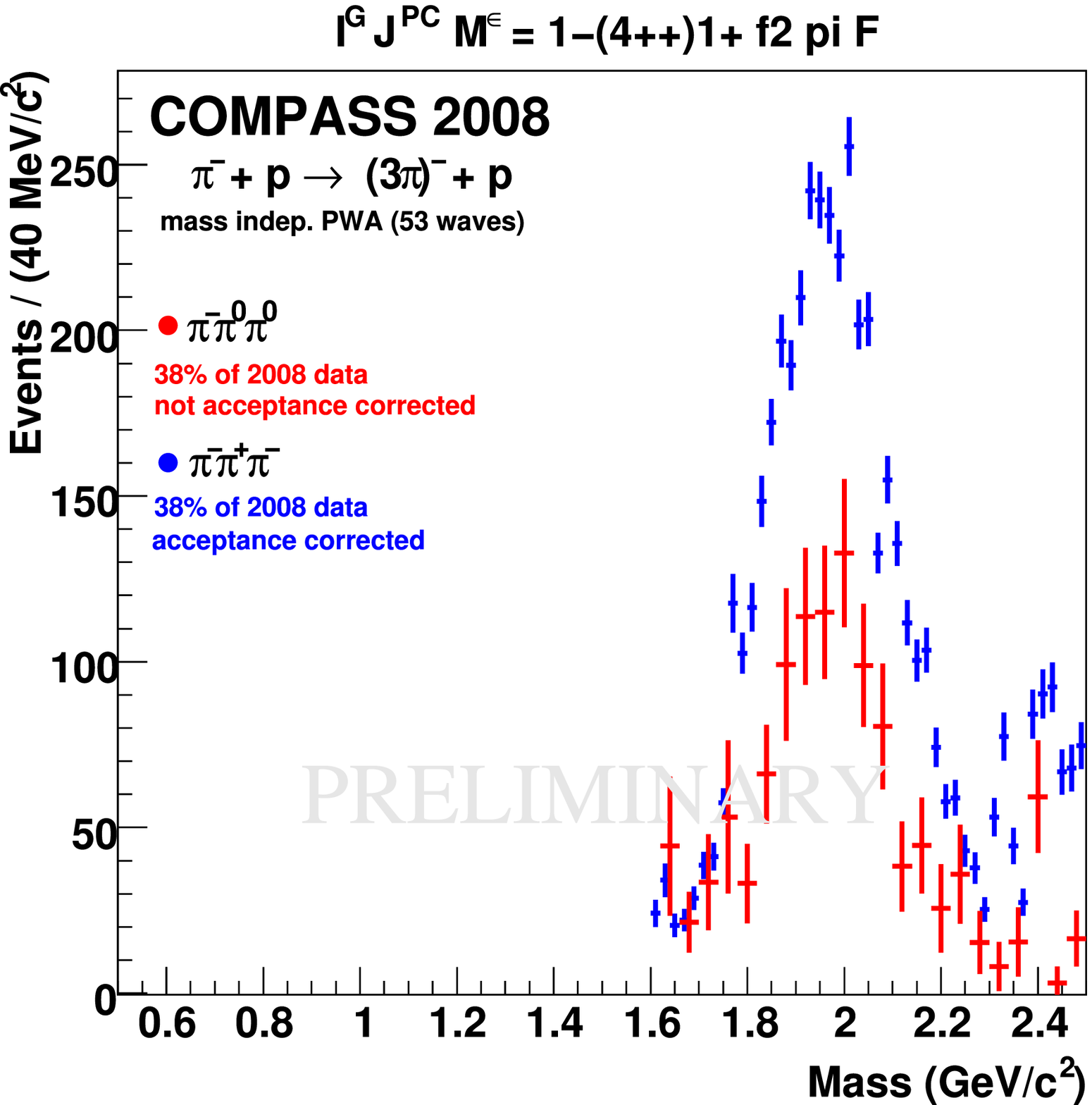} \\[-1.5ex] \centering (d)}
  \label{fig:2008_3pi_charged_vs_neutral}
  \caption{Comparison of wave intensities for \threepion\ (blue) and
    $\Ppim\Ppiz\Ppiz$ (red) final states: Intensity sum of all
    $\Pr\Ppi$ (a) and all $\Pftwo\Ppi$ waves (b). Intensity of
    \protect \wavespec{4}{+}{+}{1}{+}{\Pr\Ppi}{G} (c) and \protect
    \wavespec{4}{+}{+}{1}{+}{\Pftwo\Ppi}{F} wave (d), both with the
    $\Pafour(2040)$. (From~\cite{hadron2011_frank})}
\end{figure}

Compared to the total number of events the intensity of the
spin-exotic $1^{-+}$ wave is on the percent level. In order to extract
such small contributions reliably via PWA an excellent Monte-Carlo
description of the spectrometer acceptance is required. In this regard
the analysis of the isospin-partner final state $\Ppim\Ppiz\Ppiz$,
although having a significantly lower reconstruction efficiency, is
interesting~\cite{hadron2011_frank}. Since the reconstruction of the
\threepion\ (``charged'') and the $\Ppim\Ppiz\Ppiz$ (``neutral'')
final states relies on different parts of the apparatus, the results
can be used for internal consistency checks. By na\"ive application of
just isospin Clebsch-Gordan coefficients, one expects that an
isovector state should decay equally into charged and neutral
$(3\Ppi)^-$ states, if the decay proceeds via an isobar with isospin
1, whereas for an isoscalar isobar the intensity in the neutral decay
channel should be half of that in the charged one. This pattern is
observed in the data. Some examples are shown in
\figref{2008_3pi_charged_vs_neutral}. The two data sets are normalized
to the narrow $\Patwo(1302)$ resonance in the
\wavespec{2}{+}{+}{1}{+}{\Pr\Ppi}{D} wave. Even though no acceptance
correction was applied yet to the neutral-channel data, the intensity
sums of all $\Pr\Ppi$ waves are in good agreement, whereas the
intensity sums of all $\Pftwo\Ppi$ waves exhibit the expected
suppression factor of two in the neutral channel. This is also true
for the major waves (not shown) as well as for small-intensity waves
like the two $4^{++}$ waves shown in
\figref{2008_3pi_charged_vs_neutral}c and d. These waves contain the
$\Pafour(2040)$ resonance and illustrate the ability of COMPASS to
reconstruct resonances even in waves with percent-level intensity.

\section{$\Ppim\Petaprime$ Final State from \Ppim\ Diffraction}

\begin{figure}[b]
  \includegraphics[width=0.333\textwidth]{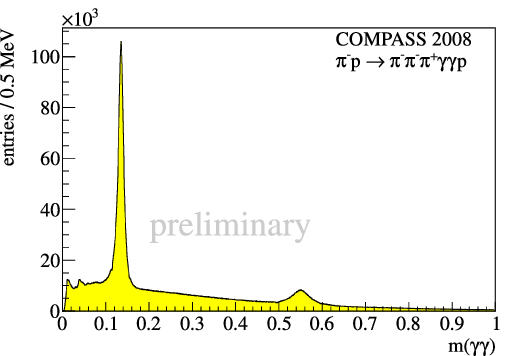}
  \includegraphics[width=0.333\textwidth]{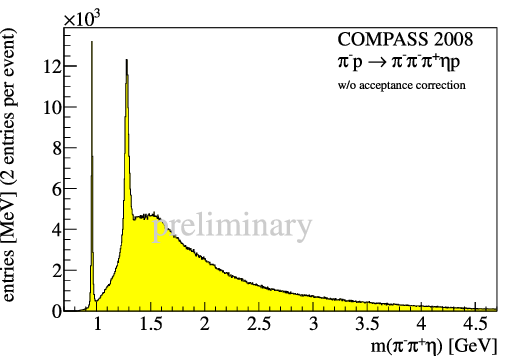}
  \includegraphics[width=0.333\textwidth]{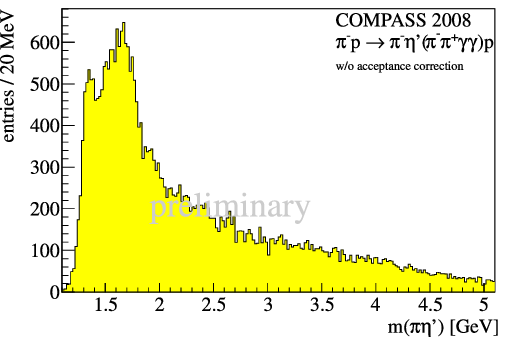}
  \label{fig:2008_etaprime_kinematics}
  \caption{Left: $\PPh\PPh$ invariant mass distribution with the
    \Ppiz\ and \Peta\ peaks for events with three charged
    tracks. Center: $\Ppip\Ppim\Peta$ invariant mass spectrum. Right:
    $\Ppim\Petaprime$ invariant mass distribution with visible
    $\Patwo(1320)$ peak. (From~\cite{hadron2011_tobias})}
\end{figure}

Other channels, where spin-exotic resonances were claimed in the past,
are $\Ppim\Peta$~\cite{eta_etaprime_pi, eta_pi_1, eta_pi_2} and
$\Ppim\Petaprime$~\cite{eta_etaprime_pi, etaprime_pi}. However, the
resonant nature of the observed signals is still
controversial~\cite{adam}. COMPASS performed an analysis of the
$\Ppim\Petaprime$ final state which is reconstructed from the decay
chain $\Petaprime \to \Ppip\Ppim\Peta$, $\Peta \to \PPh\PPh$. This is
illustrated in \figref{2008_etaprime_kinematics} which shows the
$\eta(548)$ peak in the $\PPh\PPh$ invariant mass spectrum. The
reconstructed \Peta\ are then combined with a $\Ppip\Ppim$ pair
yielding a narrow $\Petaprime(958)$ peak in the respective invariant
mass distribution. The final $\Ppim\Petaprime$ invariant mass spectrum
contains 35\,000~events and exhibits a peak from the $\Patwo(1320)$
resonance.

\begin{figure}[t]
  \includegraphics[width=\textwidth]{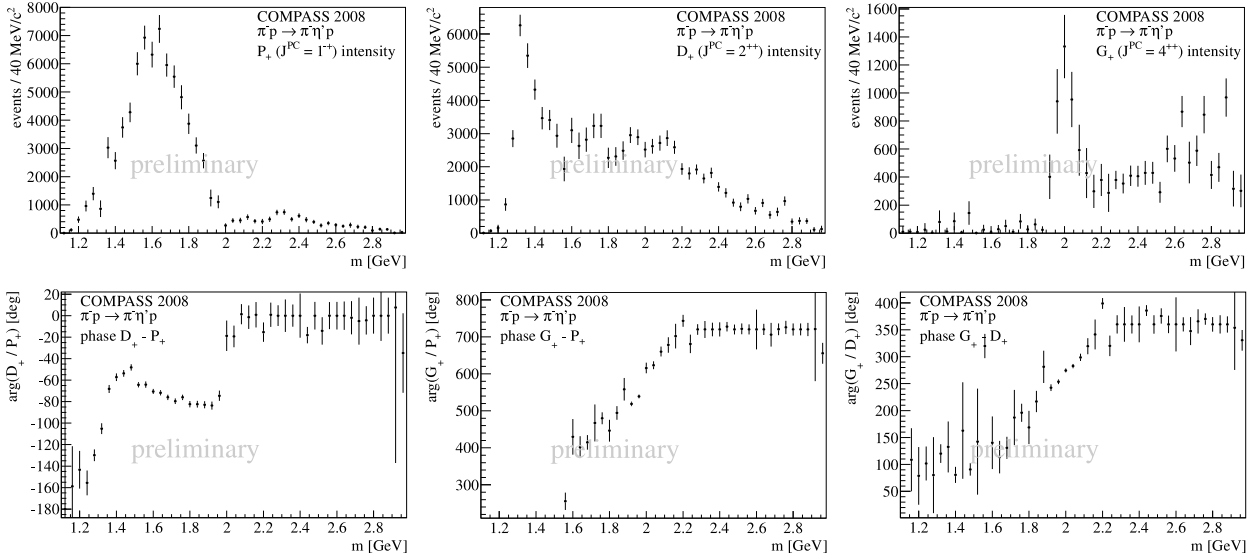}
  \label{fig:2008_etaprime_pi}
  \caption{\emph{Top row:} Intensities of major waves in
    $\Ppim\Petaprime$ final state. Spin-exotic $1^{-+}1^+$ ($P_+$)
    wave (left), $2^{++}1^+$ ($D_+$) wave with $\Patwo(1320)$
    (center), and $4^{++}1^+$ ($G_+$) wave with $\Pafour(2040)$
    (right). \emph{Bottom row:} Relative phases of major waves in
    $\Ppim\Petaprime$ final state. $D_+ - P_+$ (left), $G_+ - P_+$
    (center), and $G_+ - D_+$
    (right). (From~\cite{hadron2011_tobias})}
\end{figure}

The performed PWA~\cite{hadron2011_tobias} follows previous analyses
and includes $S$, $P$, and $D$ waves with $M \leq 1$ and both natural
and unnatural parity exchange. In addition a $4^{++}1^+$ and a
background wave were included. The top row of
\figref{2008_etaprime_pi} shows the intensities of the major
waves. The $2^{++}1^+$ ($D_+$) wave exhibits a peak of the
$\Patwo(1320)$ and the $4^{++}1^+$ ($G_+$) wave a clear signal of the
$\Pafour(2040)$. The most intense wave, however, is the spin-exotic
$1^{-+}1^+$ ($P_+$) wave which has a broad structure around 1.6\gevcc,
consistent with previous experiments. The $P_+$ wave shows slow phase
motion with respect to the $D_+$ wave in the 1.6\gevcc\ mass region
(see \figref{2008_etaprime_pi} bottom, left). A mass-dependent fit of
the spin-density matrix is work in progress.

\section{$\PKm\Ppip\Ppim$ Final State from \PKm\ Diffraction}

\begin{figure}[b]
  \includegraphics[width=0.333\textwidth]{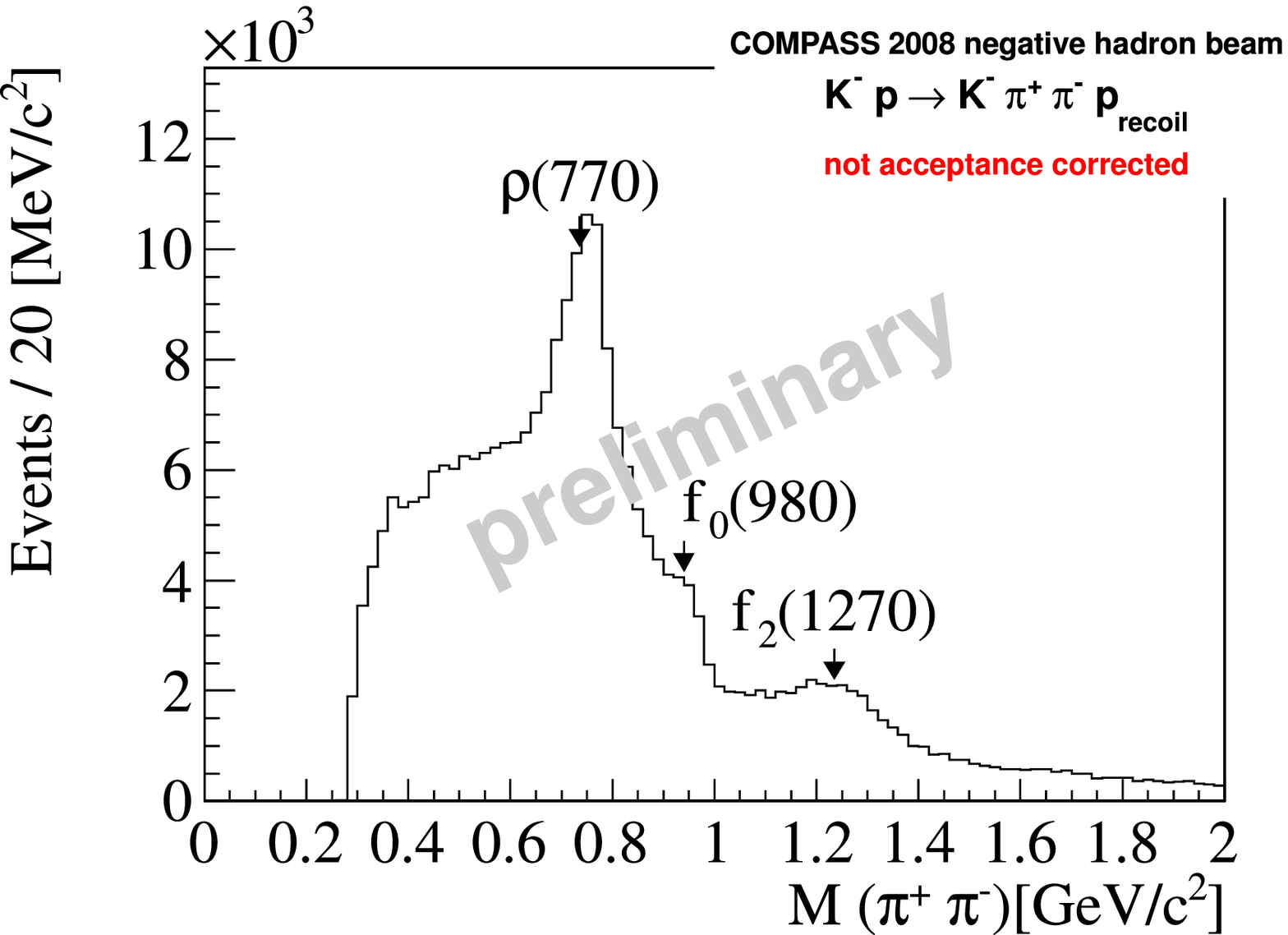}
  \includegraphics[width=0.333\textwidth]{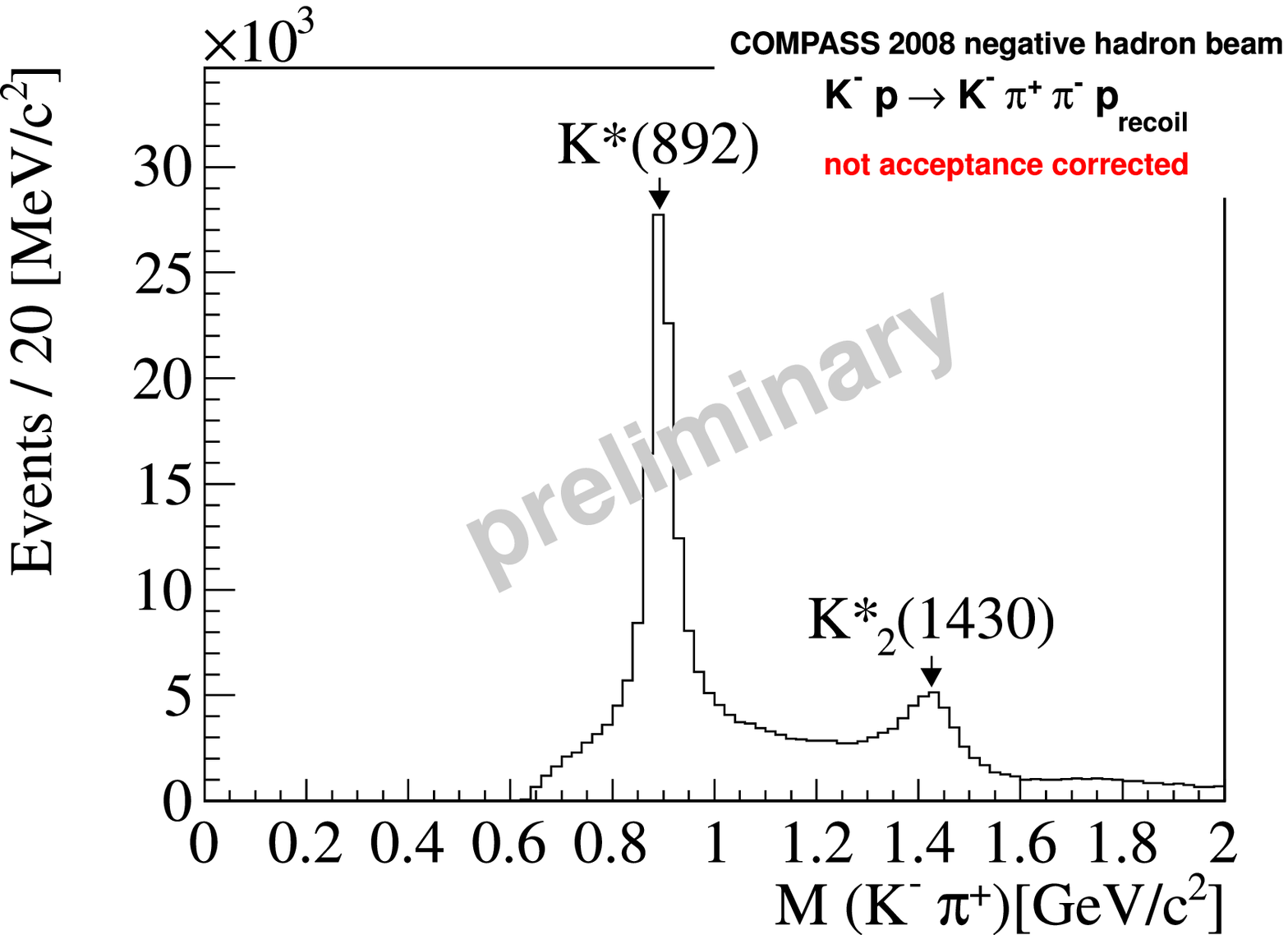}
  \includegraphics[width=0.333\textwidth]{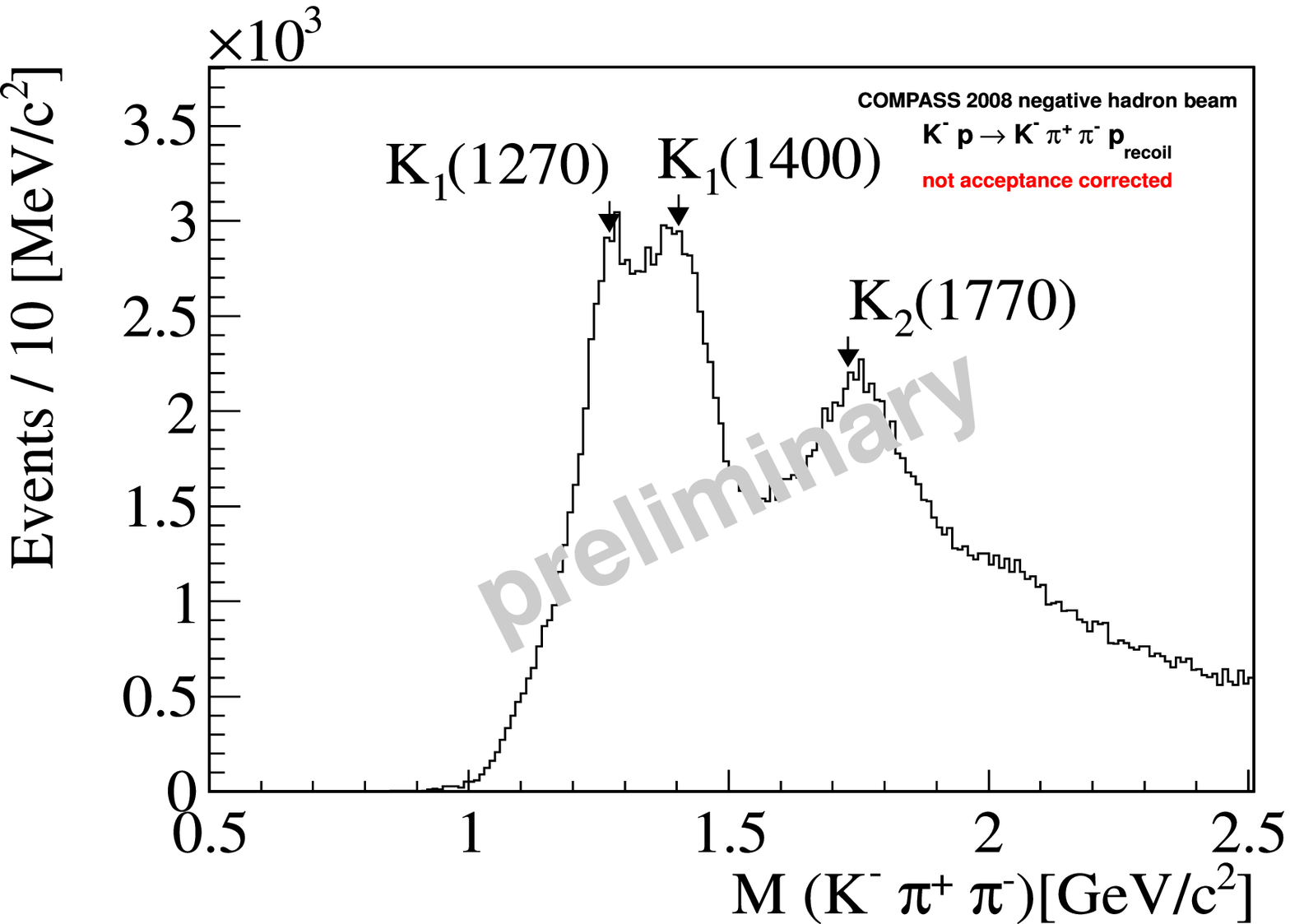}
  \label{fig:2008_kpipi_kinematics}
  \caption{Left: $\Ppip\Ppim$ invariant mass distribution for
    $\PKm\Ppip\Ppim$ events. Center: respective $\Ppip\PKm$ invariant
    mass spectrum. Right: $\PKm\Ppip\Ppim$ invariant mass
    distribution. (From~\cite{hadron2011_promme})}
\end{figure}

Although there are no spin-exotic strange mesons, because these states
are not $G$- or $C$-parity eigenstates, the spectrum of strange mesons
is still interesting as it contains many states that need
confirmation. There are also open questions about the interpretation
of some states. Since most of the available data are from the 70s and
80s, COMPASS takes the opportunity to remeasure some final states with
a state-of-the-art apparatus. An interesting channel is the
$\PKm\Ppip\Ppim$ final state which was studied by several experiments
in the past~\cite{kpipi}. The measurement~\cite{hadron2011_promme}
exploits the fact that the negative hadron beam of the M2 beam line
contains a 2.4~\% admixture of \PKm. These beam kaons were tagged by
two CEDAR detectors located 30~m upstream of the target. The
final-state kaons were identified using the RICH detector. After all
cuts a sample of 270\,000 $\PKm\Ppip\Ppim$ events was
obtained. \Figref{2008_kpipi_kinematics} shows the $\Ppip\Ppim$
invariant mass spectrum which exhibits structures from $\Pr(770)$,
$\Pfzero(980)$, and $\Pftwo(1270)$. The corresponding $\Ppip\PKm$
spectrum shows peaks from $\PKstar(892)$ and $\PKtwostar(1430)$. Also
the $m_{\PKm\Ppip\Ppim}$ distribution shows significant structures.

\begin{figure}
  \includegraphics[width=0.333\textwidth]{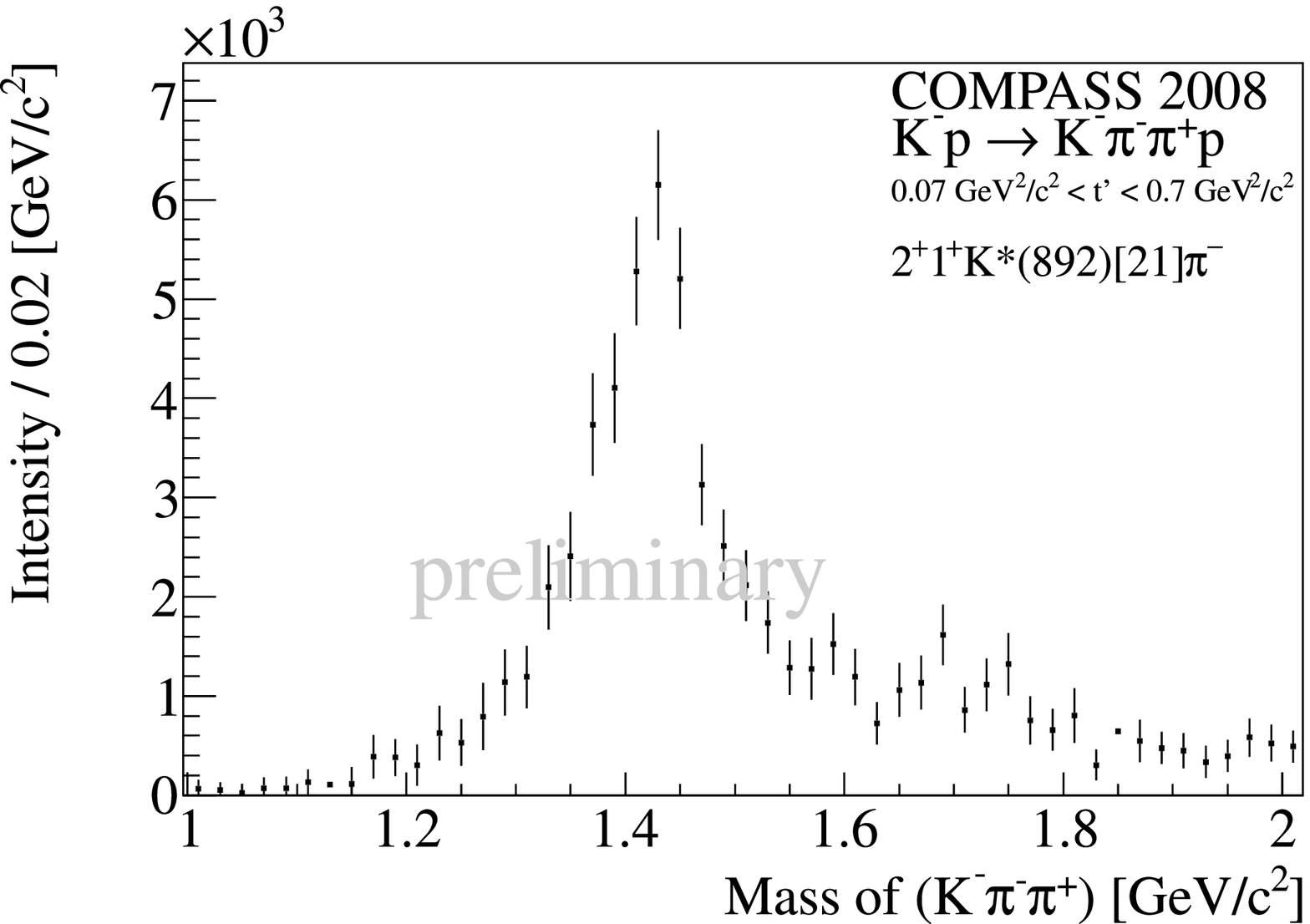}
  \includegraphics[width=0.333\textwidth]{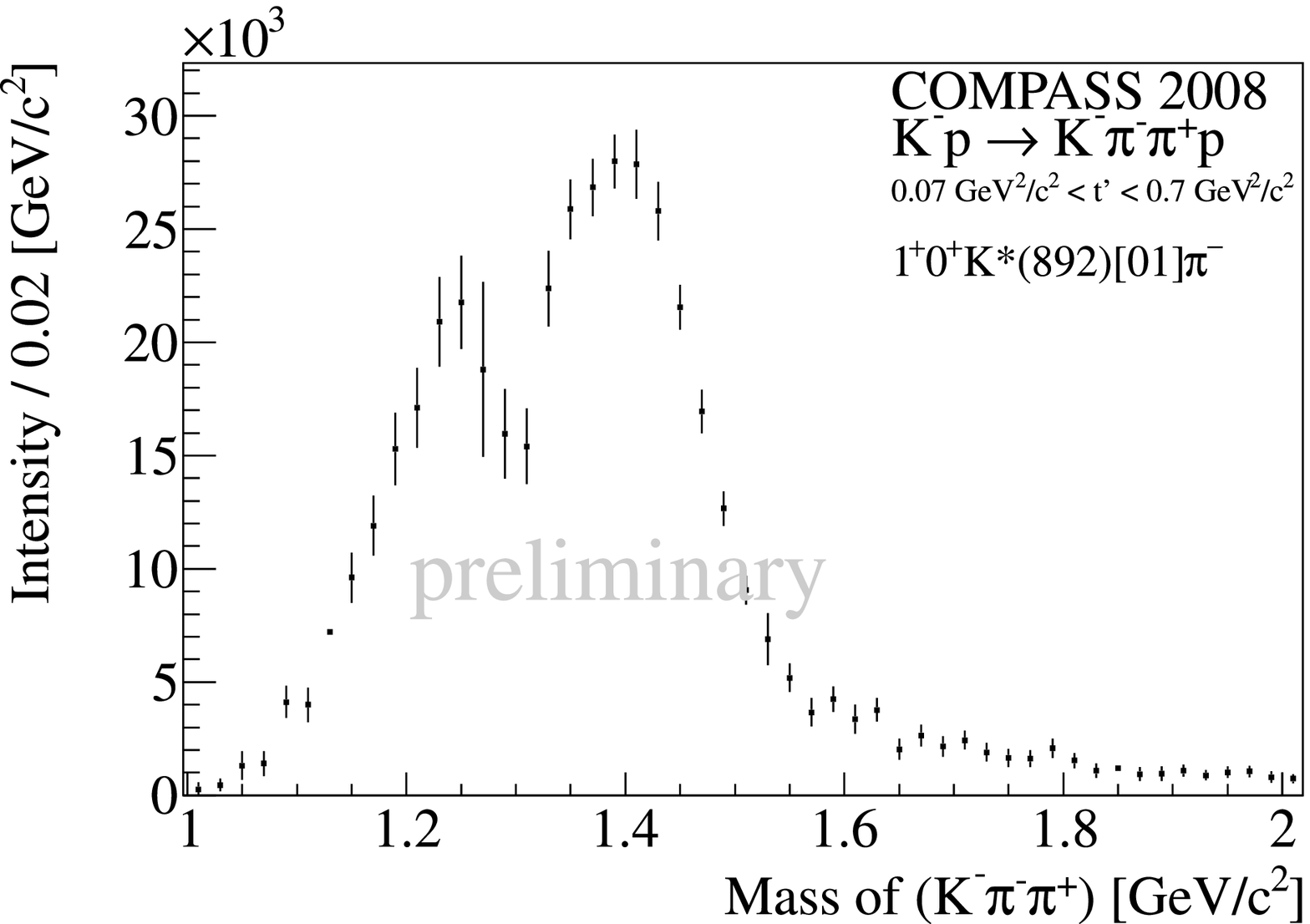}
  \includegraphics[width=0.333\textwidth]{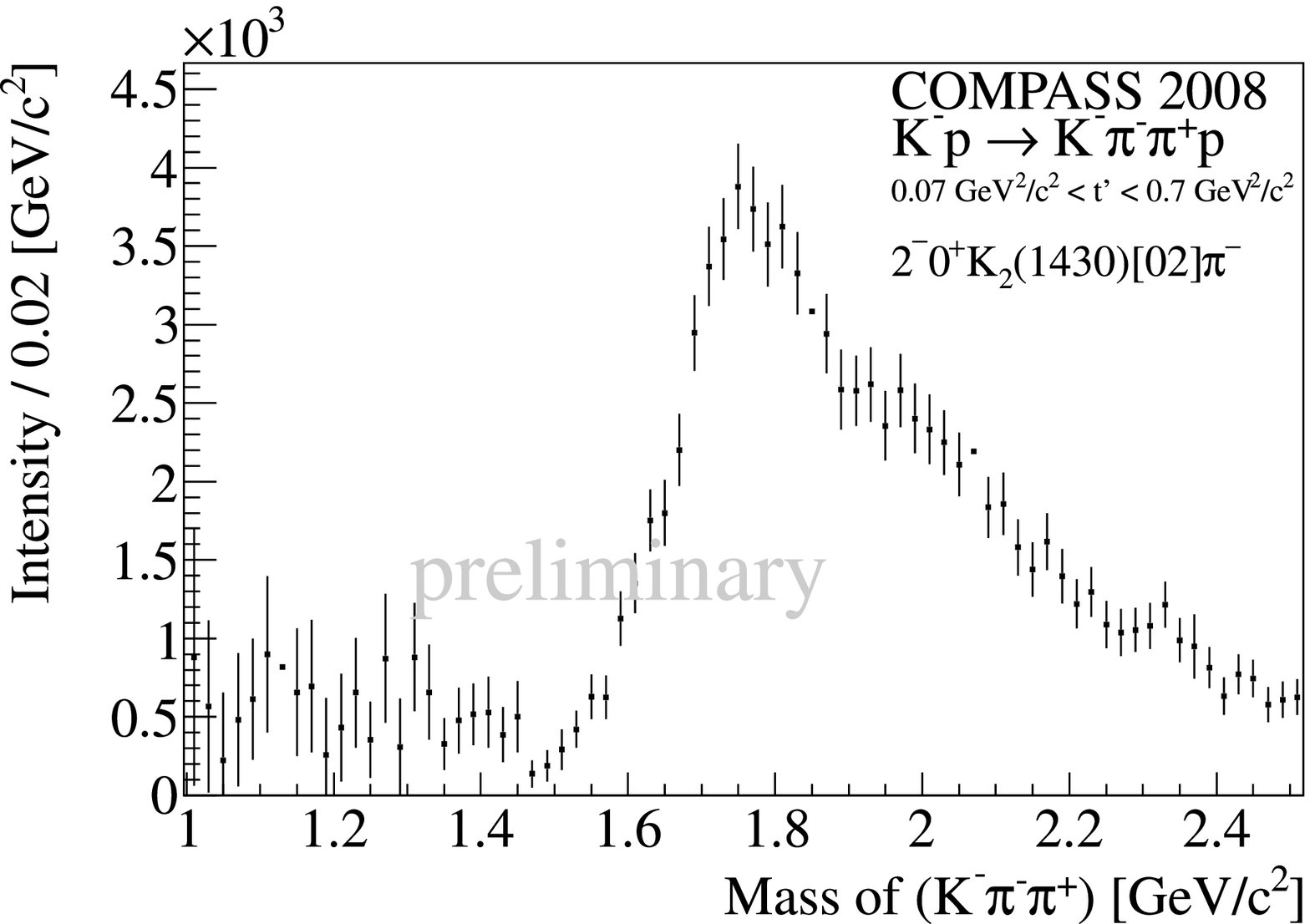}
  \label{fig:2008_kpipi_major_waves}
  \caption{Intensities of some dominant waves in the $\PKm\Ppip\Ppim$
    final state: \protect \wavespec{2}{+}{}{1}{+}{\PKstar(892)\Ppi}{D}
    wave (left), \protect \wavespec{1}{+}{}{0}{+}{\PKstar(892)\Ppi}{S}
    wave (center), and \protect
    \wavespec{2}{-}{}{0}{+}{\PKtwostar(1430)\Ppi}{S} wave
    (right). (From~\cite{hadron2011_promme})}
\end{figure}

A PWA using a model with 19~waves plus a background wave together with
a rank-2 spin-density matrix was performed in 20\mevcc\ wide mass
bins. Some of the most prominent waves are shown in
\figref{2008_kpipi_major_waves}. A clear $\PKtwostar(1430)$ peak is
seen in the \wavespec{2}{+}{}{1}{+}{\PKstar(892)\Ppi}{D} wave. The
\wavespec{1}{+}{}{0}{+}{\PKstar(892)\Ppi}{S} wave shows two peaks that
probably belong to the $\PKone(1270)$ and the $\PKone(1400)$. As
expected the $\PKone(1400)$ peak is absent in the corresponding
\wavespec{1}{+}{}{0}{+}{\Pr\PK}{S} wave (not shown). Similarly the
broad structure in the
\wavespec{2}{-}{}{0}{+}{\PKtwostar(1430)\Ppi}{S} wave, which could be
due to $\PKtwo(1770)$ and $\PKtwo(1820)$, becomes narrower in the
\wavespec{2}{-}{}{0}{+}{\Pftwo(1270)\PK}{S} wave (not shown). More
information on these and other states will be extracted by fitting the
mass dependence of the spin-density matrix.

\section{Conclusions}

First partial-wave analyses of the large data set of diffractive
dissociation events that COMPASS has collected show interesting
results in various channels. Structures around 1.6\gevcc\ are observed
in spin-exotic $\jpc = 1^{-+}$ waves in the $\Pr\Ppi$ and the
$\Ppi\Petaprime$ decay channels. The resonant nature of these bumps,
however, still has to be verified by mass-dependent fits. The same is
true for some structures seen in the $\PKm\Ppip\Ppim$ final
state. With further improved analyses COMPASS will make a significant
contribution to the study of the light-quark meson spectrum.

%%%%%%%%%%%%%%%%%%%%%%%%%%%%%%%%%%%%%%%%%%%%%%%%
%% BACKMATTER
%%%%%%%%%%%%%%%%%%%%%%%%%%%%%%%%%%%%%%%%%%%%%%%%

\begin{theacknowledgments}
  This work is supported by the German BMBF, the Maier-Leibnitz-Labor
  der LMU und TU M\"unchen, the DFG Cluster of Excellence \emph{Origin
    and Structure of the Universe}, and CERN-RFBR grant 08-02-91009.
\end{theacknowledgments}

\bibliographystyle{aipproc}   % if natbib is available

\end{document}